\documentclass{article}

\usepackage[dvipsnames,svgnames,table]{xcolor}
\usepackage{amssymb, amsmath, verbatim, epsfig}
\usepackage[sort&compress,comma,super]{natbib}
\usepackage{amssymb,verbatim,epsfig}
\usepackage[margin=1in]{geometry} 
\usepackage[capitalize]{cleveref}
\usepackage[small,bf]{caption}
\usepackage{amsmath,setspace}
\usepackage[utf8]{inputenc}
\usepackage{siunitx}
\usepackage{multicol}
\usepackage{multirow}
\usepackage{setspace}
\usepackage{arydshln}
\usepackage{textcomp}
\usepackage{mathrsfs}
\usepackage{lineno}
\usepackage{lipsum}
\usepackage{url}

\newcommand{\volume}{{\ooalign{\hfil$V$\hfil\cr\kern0.08em--\hfil\cr}}}

\makeatletter
\newcommand{\thickhline}{
    \noalign {\ifnum 0=`}\fi \hrule height 2pt
    \futurelet \reserved@a \@xhline
}
\newcolumntype{"}{@{\hskip\tabcolsep\vrule width 2pt\hskip\tabcolsep}}
\makeatother

\title{\bf 
\vspace{-0.75in}
Droplet impact of emulsions on inclined microstructured surfaces}

\author{Ulises Rosales-Gallegos$^1$, Laura Oropeza-Ramos$^2$, Francisco Manuel Sánchez-Arévalo$^1$ \\ and Miguel A. Quetzeri-Santiago$^{1*}$ \\
 \small $^\text{1}$Instituto de Investigaciones en Materiales, Universidad Nacional Aut\'onoma de M\'exico, \\ \small 04510, Cd. Universitaria, Mexico City, Mexico. \\
 \small $^2$ Facultad de Ingeniería, Universidad Nacional Autónoma de México, 04510, Cd. Universitaria Mexico City, Mexico.\\
 \small $^*$ email: mquetzeri@materiales.unam.mx}
 \date{}

\begin{document}
\maketitle

\begin{abstract}
Droplet impacts rarely occur on horizontal smooth surfaces, yet the combined influence of surface inclination, microstructure, and fluid composition remains poorly understood for emulsions. Here, we experimentally investigate the impact of 20\% (v/v) silicone oil-in-water emulsion droplets on smooth glass and micropillar arrays with two interpillar spacings (50 and 100 $\mu$m) inclined at 55$^\circ$. The role of surface wettability is further examined using a superhydrophobic coating. High-speed imaging is employed to characterise the impact dynamics over a wide range of Weber numbers ($18 \leq We \leq 565$), while optical microscopy is used to quantify the residual wetted area after impact. Five impact regimes are identified: deposition, sliding, partial rebound, complete rebound, and splashing. Surface inclination introduces asymmetry in spreading and recoil, whereas microstructured substrates reduce adhesion and promote rebound by decreasing the liquid-solid contact area. In contrast to water, the emulsion exhibit enhanced deposition on smooth surfaces, increased sliding on uncoated microstructures, and earlier ligament formation and secondary breakup at high Weber numbers. The dispersed oil phase locally modifies wetting and lubrication, producing liquid bridges, heterogeneous recoil, and larger residual wetted areas on smooth substrates. On microstructured surfaces, however, capillary confinement within the texture limits the projected wetted area despite increased liquid retention inside the microstructure. These results demonstrate that the impact behaviour of emulsions cannot be predicted solely from conventional dimensionless groups, as local phase redistribution fundamentally alters energy dissipation, wetting, and fragmentation. The findings provide new insight into spray deposition on inclined textured surfaces relevant to agricultural spraying and coating technologies.
\end{abstract}
\vspace{0.1 in}
\noindent{\bf Keywords:} [droplet impact, splashing, wettability, emulsions]

\section{Introduction}
The dynamics of droplet impact on solid surfaces is a fundamental phenomenon central to numerous industrial and environmental processes, including inkjet printing, spray cooling, and agricultural pesticide application \cite{damak2022dynamics, xu2024impact, abbot2025elucidating}. The outcome of such an impact, i.e., whether the droplet spreads, retracts, rebounds, or splashes, is governed by a complex interplay of inertial, viscous, capillary, and adhesion forces \cite{yarin2006drop, josserand2016drop}. These dynamics are typically characterised using dimensionless parameters such as the Weber number ($We = \rho D_0 U^2/\gamma$), representing the ratio of inertia (droplet density $\rho$, diameter $D_0$ impact velocity $U$) to surface tension ($\gamma$), and the Reynolds number ($Re = \rho D_0 U/\mu$), relating inertia to viscous forces represented by the viscosity of the liquid $\mu$ \cite{yarin2006drop, josserand2016drop}. While these parameters effectively model single-phase Newtonian fluids, they are often insufficient for describing the behaviour of complex mixtures like emulsions, where the presence of a dispersed phase introduces intricate rheological and interfacial interactions \cite{quetzeri2024droplet, abbot2025elucidating}.

Emulsions, consisting of droplets of one immiscible liquid dispersed within another continuous phase, are widely encountered in food processing, cosmetics, pharmaceuticals, agrochemical sprays, coating technologies, and enhanced oil recovery \cite{mcclements2015food, tadros2013emulsion, schramm2005emulsions}. Their technological relevance stems from the ability to combine otherwise incompatible liquids and tailor bulk viscosity, stability, lubrication, and interfacial properties through droplet size distribution, dispersed-phase fraction, and surfactant formulation \cite{bibette1999emulsion, mcclements2015food}. Under dynamic conditions such as spraying, jetting, or impact, these additional degrees of freedom can strongly modify spreading, recoiling, splashing, and deposition compared with single-phase liquids, since interfacial deformation, droplet coalescence, and phase redistribution may occur during the very short impact timescale \cite{damak2022dynamics, piskunov2025spreading, quetzeri2024droplet}. Despite this practical importance, the vast majority of fundamental studies on emulsion impact have considered normal incidence onto horizontal substrates, where the impact symmetry simplifies both experimentation and modelling. In contrast, many real operating conditions involve droplets striking surfaces at an angle, such as on turbine blades, aircraft wings, solar panels, or plant leaves, where inclination introduces tangential momentum and directional asymmetry that can substantially alter impact outcomes \cite{vsikalo2005impact,wang2021dynamics, voytkova2024dynamics}.

Impacts on inclined substrates introduce a critical directional asymmetry \cite{vsikalo2005impact, krishna2026droplet, bae2024bouncing, saha2025experimental, yu2024symmetry}. Unlike isotropic spreading on horizontal surfaces, an oblique impact decomposes the initial momentum into normal and tangential components, leading to elongated droplet shapes and shifted splashing thresholds \cite{xu2022droplet,saha2025experimental}. This surface inclination significantly influences the contact time, maximum spreading diameter, and the transition between different impact regimes \cite{bae2024bouncing, chowdhury2026effect, sahoo2019interplay}. On microstructured surfaces, this inclination can further exacerbate the asymmetry of spreading and retraction, as the tangential inertial force drives sliding while the microstructures provide anisotropic pinning or enhanced lubrication through trapped air pockets \cite{shi2025aerodynamic, cao2025dual, qian2025experimental}.

In this work, we extend our previous investigation \cite{quetzeri2024droplet} to study the impact of silicone oil-in-water emulsions (20\% v/v) on both smooth and microstructured surfaces with different wettabilities at a fixed inclination angle of 55° with respect to the horizontal. Using high-speed imaging in a shadowgraph configuration, we analyse how surface morphology and inclination define the preferred direction of motion and modify the competition between inertial, viscous, and capillary forces. We characterise the physical significance of the Weber number across different regimes. Furthermore, we explore how the oil phase fundamentally alters impact behaviour by promoting pinning or enhancing mobility depending on the Weber number, ultimately breaking the isotropic symmetry characteristic of horizontal impacts.

\section{Methods}

\subsection{Emulsion preparation}

Two liquids were used throughout the experiments: water (Milli-Q Advantage A10) and silicone oil with a nominal kinematic viscosity of 50 cSt (Kausil, Caucho Químico).

The emulsion was prepared by mixing 28.8 mL of water with 7.2 mL of silicone oil in a 40 mL glass vial using a micropipette (Sartorius Tacta 5000). The mixture was then emulsified using an ultrasonic probe (DW-Technologies) immersed in the liquid. Pulses of 2 s duration were applied at a frequency of 24.6 kHz at 800 W, with 1 s resting intervals between pulses, for a total processing time of 7 minutes.

The resulting liquid was a stable white oil-in-water emulsion with an average dispersed droplet diameter of approximately 5 $\mu$m \cite{quetzeri2024droplet}. The physical properties of the tested fluids are summarised in Table \ref{tab:fluids}. The emulsion was subsequently transferred into a 20 mL glass syringe for the droplet impact experiments.

\begin{table}[h!]
\centering
\caption{Physical properties of the fluids used in this study at 25$^\circ$C. $\mu$ is the dynamic viscosity, $\gamma$ is the surface tension, $\rho$ is the density, and $\sigma$ is the interfacial tension between silicone oil and water \cite{quetzeri2024droplet}.}
\label{tab:fluids}
\begin{tabular}{lcccc}
\hline
Fluid & $\mu$ (mPa$\cdot$s) & $\gamma$ (mN/m) & $\rho$ (kg/m$^3$) & $\sigma$ (mN/m) \\
\hline
Water & 0.9 & 72.8 & 997.0 & -- \\
Silicone oil (50 cSt) & 4.6 & 19.7 & 913.0 & 25.0 \\
Emulsion & 15.4 & 67.1 & 980.2 & -- \\
\hline
\end{tabular}
\end{table}

\subsection{Droplet generation and impact experiments}

Droplet impact experiments were conducted on three different surfaces inclined at 55$^\circ$ with respect to the horizontal.

\begin{figure}[h]
    \centering
    \includegraphics[width=0.9\linewidth]{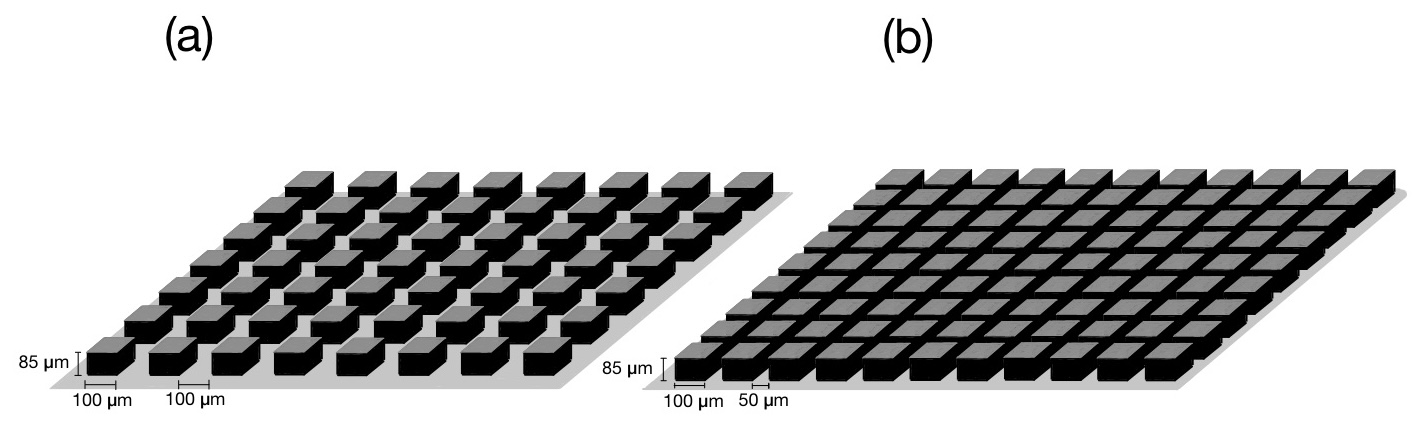}
    \caption{Schematic representation of the silicon substrates patterned with pillar-type microstructures: (a) surface M50 with an inter-pillar spacing $L$ of 50 $\mu$m, and (b) surface M100 with an inter-pillar spacing $L$ of 100 $\mu$m.}
    \label{fig:diagramasuperficies}
\end{figure}

The substrates consisted of a smooth glass slide (Lauka) and two silicon wafers patterned with pillar-type microstructures. Both microstructured surfaces had pillars of 85 $\mu$m height and 100 $\mu$m width. The difference between them was the inter-pillar spacing $L$, which was 50 $\mu$m for the first surface and 100 $\mu$m for the second (see Fig. \ref{fig:diagramasuperficies}). These surfaces are hereafter referred to as M50 and M100, respectively.

The microstructures were fabricated using SU-8 3035 photoresist spin-coated onto silicon wafers following the manufacturer recommendations. The CAD-designed pillar pattern was transferred onto the polymer layer by UV photolithography. After exposure, the uncrosslinked regions were removed during development, while the exposed regions remained attached to the substrate. Pre- and post-exposure baking steps were performed for 5 min at 95$^\circ$C. The developed surfaces were rinsed with isopropanol, followed by a final hard-bake at 200$^\circ$C for 30 min to improve mechanical robustness and chemical resistance \cite{quetzeri2024droplet}.

\begin{figure}
    \centering
    \includegraphics[width=0.5\linewidth]{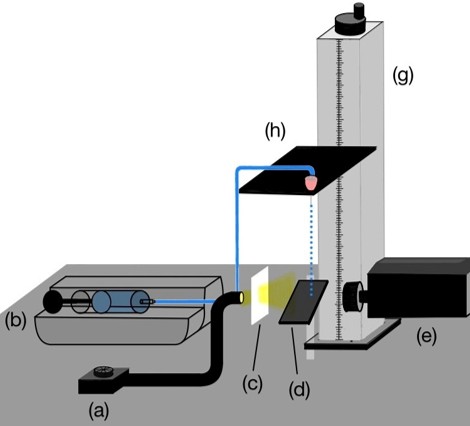}
    \caption{Schematic of the experimental setup used for droplet generation and high-speed visualisation of impacts on inclined surfaces. The main components are: (a) DiCon LED fiber-optic light source, (b) SinoHero 508Hi syringe pump, (c) light diffuser, (d) inclined test surface, (e) Phantom v9.1 high-speed camera, (g) vertically adjustable support column, and (h) base platform.}
    \label{fig:diagrama1}
\end{figure}

The experimental setup was mounted on an optical table (TMC). It consisted of an inclined sample holder fixed at 55$^\circ$, and a 21 G needle connected to a syringe pump (SinoHero 508Hi) mounted on a vertically adjustable column for droplet generation. Impact dynamics were recorded using a Phantom v9.1 high-speed camera in a shadowgraph configuration, illuminated by a fiber-optic LED light source through a diffuser (see Fig. \ref{fig:diagrama1}).

Experiments were performed using both water droplets and emulsion droplets. The impact velocity was controlled by varying the release height of the needle from 20.6 cm to 71.6 cm, resulting in velocities within the range $U_0 = 0.697$--$3.472$ m s$^{-1}$ and Weber numbers $We = 17.75$--$565.45$. This range enabled the observation of several impact outcomes, including deposition, sliding, rebound, and splashing. At least three independent repetitions were performed for each height, surface, and liquid.

To investigate the role of wettability, each surface was coated with a commercial superhydrophobic treatment (Glaco). The coating was applied in two layers with a drying time of 1 h between applications. After each experiment, the surface was cleaned with toluene to remove the coating and allowed to dry completely before reapplication for the next trial.

For imaging purposes, the camera was initially aligned perpendicular to the optical table so that the droplet trajectory appeared vertical in the recorded videos. For this videos the camera was set at 779 fps with and exposure time of 70 $\mu$s. However, at higher impact velocities, the post-impact motion extended beyond the field of view. To overcome this limitation, the camera was rotated until its optical axis was normal to the inclined surface, thereby increasing the observable area around the impact region. For these experiments the camera was set at 1500 fps with an exposure time of 100 $\mu$s.

\subsection{Image analysis and measurement procedures}

Droplet diameter and impact velocity were obtained from the high-speed videos using a custom image-processing routine developed in MATLAB. Each frame was first binarised, and the first frame was subtracted from the subsequent frames to isolate the droplet from the background. The droplet contour was then detected, allowing its diameter to be measured in pixels.

The horizontal and vertical positions of the droplet were determined from the rightmost and bottommost pixels of the detected contour, respectively. Using the calibrated pixel size and frame rate, the instantaneous velocity components $U_x$ and $U_y$ were calculated. For videos recorded with the rotated camera configuration, the droplet trajectory appeared diagonal in the image plane. In these cases, the impact velocity was determined from the magnitude of the velocity vector, $U_0 = \sqrt{U_x^2 + U_y^2}$.

The wetted area remaining on the surface after impact was measured using an optical microscope (Zeiss AX10 Scope A1). Videos of the deposited stain were recorded through the microscope camera. In cases where the stain exceeded the field of view, the full image was reconstructed from multiple overlapping photographs with specialised software. The wetted region was identified by pixel selection, and the total number of pixels was converted into physical area using a calibrated spatial scale.

\section{Results and discussion}
This section summarises the impact dynamics of water and emulsion droplets on smooth glass and the microstructured surfaces M50 and M100. High-speed imaging was used to identify the main impact regimes and to quantify the influence of substrate morphology, wettability, and liquid properties.

\subsection{Observed impact regimes}

\begin{figure}[htbp]
        \centering
        \includegraphics[width=0.8\textwidth]{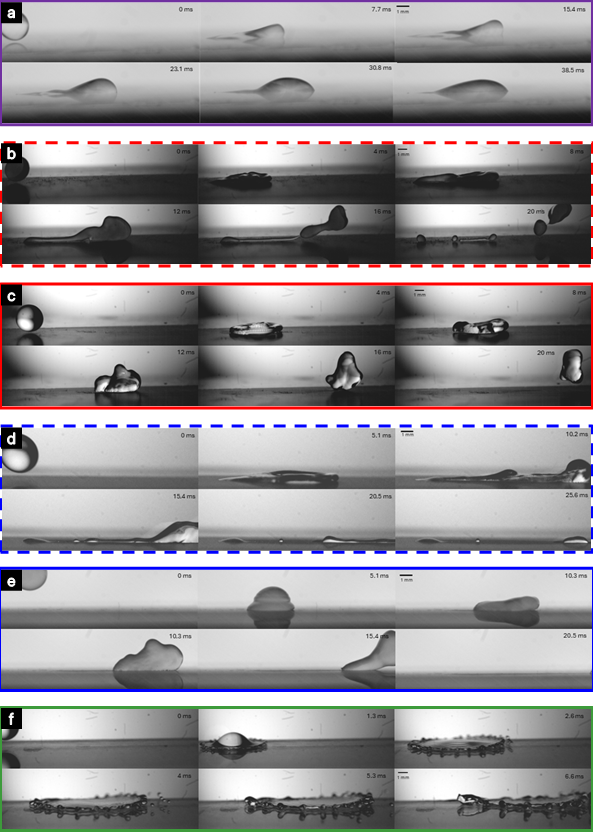}
        \caption{Representative impact regimes observed during emulsion droplet impact on an inclined surface: (a) deposition, where the droplet remains attached to the substrate after damped oscillations; (b) partial rebound, in which a portion of the droplet detaches while a residual liquid fraction remains pinned to the surface; (c) complete rebound, characterised by full retraction and detachment from the substrate; (d) partial sliding, where the droplet translates along the inclined surface while a fraction of the liquid remains adhered; (e) sliding, in which the droplet maintains continuous contact with the substrate while moving downslope; and (f) splashing, characterised by the formation of liquid fingers and ligaments that subsequently fragment into secondary droplets.}
    \label{fig:regimenes}
\end{figure}

Figure \ref{fig:regimenes} summarises the five impact regimes identified in the present study: deposition, partial rebound, complete rebound, sliding, and splashing. These regimes are consistent with previous observations of droplets impacting inclined surfaces \cite{qian2025experimental,fan2026impact,vsikalo2005impact}. However, the presence of emulsions and surface microstructures modifies the Weber number at which each regime emerges, as discussed in the following sections. In addition, surface inclination introduces a pronounced asymmetry in the impact dynamics that is absent in normal impacts. Immediately after contact, the droplet flattens into an elongated lamella whose major axis aligns with the tangential component of the impact velocity. Consequently, the downstream contact line advances more rapidly than the upstream one, producing an asymmetric liquid sheet that subsequently retracts or fragments depending on the balance between inertia, capillarity, viscous dissipation, and adhesion.

In the deposition regime (Fig.~\ref{fig:regimenes}a), the expanding lamella loses sufficient kinetic energy through viscous dissipation and substrate adhesion that capillary retraction cannot overcome the adhesive forces. The droplet undergoes several damped oscillations before reaching its equilibrium sessile shape, remaining permanently attached to the surface.

Partial rebound (Fig.~\ref{fig:regimenes}b) occurs when capillary forces are sufficient to detach only part of the liquid volume. During retraction, the downstream portion of the droplet lifts from the substrate, while a residual liquid bridge remains anchored near the impact point. This bridge eventually ruptures, leaving a small volume of liquid deposited on the surface while the remaining fluid rebounds.

Complete rebound (Fig.~\ref{fig:regimenes}c) is observed when adhesion and viscous losses are sufficiently reduced for the entire droplet to retract and detach from the substrate. The droplet first recoils into a compact shape before leaving the surface predominantly in the direction normal to the substrate, with only a small tangential displacement induced by the inclined impact.

The sliding regime (Fig.~\ref{fig:regimenes}d--e) is characterised by sustained contact between the droplet and the substrate throughout the impact event. Following the initial asymmetric spreading, the droplet translates downstream driven by the tangential component of inertia and gravity while continuously deforming into an elongated shape. Depending on the surface properties and impact velocity, the droplet may maintain a rounded cap (Fig.~\ref{fig:regimenes}e) or develop a long, thin trailing film (Fig.~\ref{fig:regimenes}d) as viscous stresses compete with capillary contraction.

Splashing (Fig.~\ref{fig:regimenes}f) occurs at the highest Weber numbers used in this study, where inertial forces dominate over capillary stabilization. The rapidly expanding lamella develops pronounced fingers around its rim that subsequently stretch into ligaments and fragment into multiple secondary droplets. On inclined surfaces, the splashing process is asymmetric, with the majority of the ejected droplets originating from the downstream edge where the tangential momentum promotes faster lamella expansion and destabilization.



\subsubsection{Effect of surface properties}

\begin{figure}[htbp]
        \centering
        \includegraphics[width=1\textwidth]{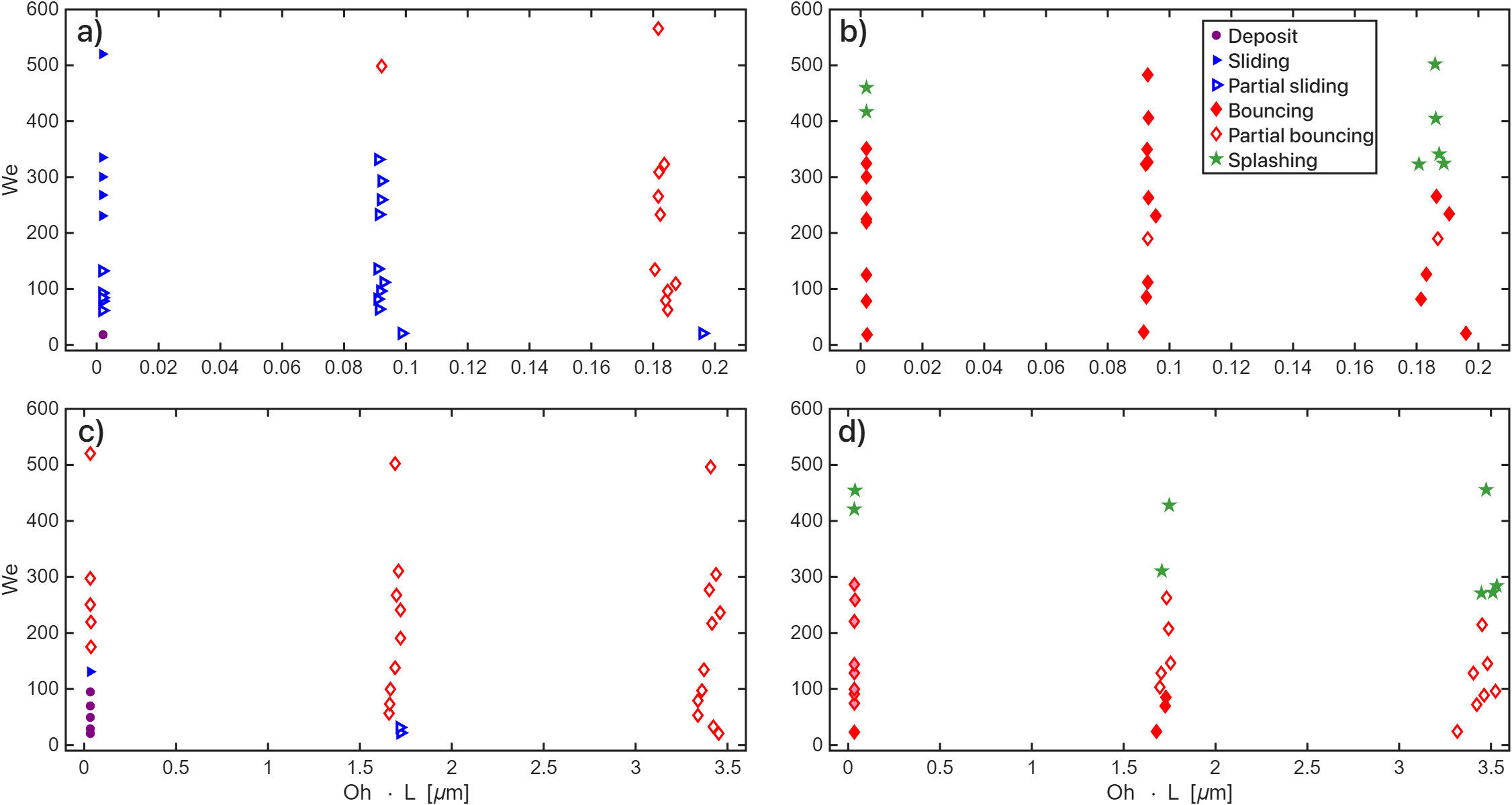}
        \caption{Impact regime maps for droplets impacting inclined surfaces as a function of Weber number ($We$) and the product of the Ohnesorge number with the inter-pillar spacing ($Oh \cdot L$) . Results are shown for (a) water on uncoated substrates, (b) water on superhydrophobic (Glaco-coated) substrates, (c) silicone oil-in-water emulsion on uncoated substrates, and (d) silicone oil-in-water emulsion on Glaco-coated substrates. The symbols identify the observed impact outcomes: deposition, sliding, partial rebound, complete rebound, and splashing. Surface morphology and wettability strongly modify the transitions between regimes, while the emulsion exhibits distinct impact behaviour compared with water, particularly on microstructured substrates.}
    \label{fig:diagramas}
\end{figure}

For uncoated surfaces, glass mainly promoted deposition or sliding, while microstructured substrates favoured partial rebound (Fig.~\ref{fig:diagramas} a). The trapped air between pillars reduced the liquid--solid contact area and lowered adhesion, facilitating droplet recoil and mobility \cite{bartolo2006bouncing,yao2020sliding}. Deposition was only observed on glass at low Weber number ($We \approx 18$), whereas M50 and M100 already showed sliding at comparable conditions.

Applying the superhydrophobic coating (Glaco) strongly altered the dynamics (see Fig.~\ref{fig:diagramas} b) . Complete rebound became the dominant regime on all surfaces from $We \approx 20$, indicating reduced viscous dissipation and more efficient restitution of kinetic energy. Similar behaviour has been reported for water impacts on coated microstructured surfaces under normal incidence \cite{dwivedi2024dynamics, xia2026droplet}.

At high Weber numbers, splashing was absent on uncoated surfaces but emerged on coated glass and M100 for $We \approx 320$--420. This suggests that reducing adhesion lowers the splashing threshold, in agreement with previous studies linking splashing to contact angle effects \cite{quintero2019splashing, quetzeri2019role}. On coated microstructured surfaces, an intermediate partial rebound regime was also observed near $We \approx 190$, where inertia and hydrophobicity locally balanced, leaving micron-sized residual droplets.

The emulsion exhibited markedly different behaviour from water (see Fig.~\ref{fig:diagramas} c) and d). On uncoated glass, deposition persisted up to $We \leq 100$, reflecting the larger viscosity of the emulsion and enhanced energy dissipation. On M50 and M100, however, microstructures promoted loss of contact and partial rebound even at Weber numbers as low as 50.

For $We \approx 20$, M50 showed sliding, whereas M100 displayed partial rebound. Increasing pillar spacing therefore enhanced mobility and detachment, likely due to reduced pinning and lower viscous resistance inside the microtexture.

After Glaco coating, rebound in the normal direction increased substantially (see Fig.~\ref{fig:diagramas} d). Coated glass and M50 showed complete rebound at $We \approx 20$, whereas coated M100 still exhibited localised penetration and partial rebound. This indicates that increased hydrophobicity does not always suppress liquid penetration under dynamic impact; rather, reduced dissipation can increase local pressure and drive transient impregnation of the texture \cite{xia2026droplet}.

At higher Weber numbers, coated surfaces promoted splashing. On coated glass, crown-like fingers formed and detached at $We \approx 420$. At $We \approx 300$ on M50 and M100, the emulsion often produced anchored liquid residues connected to the rebounding drop by unstable filaments that later fragmented. The transition to splashing occurred at lower Weber number for M100 than for M50, showing that larger $L$ destabilises the expanding lamella and facilitates breakup \cite{zhang2018tunable, quetzeri2024droplet, iqbal2025droplet}.

\subsubsection{Effect of liquid properties}

\begin{figure}[htbp]
        \centering
        \includegraphics[width=1\textwidth]{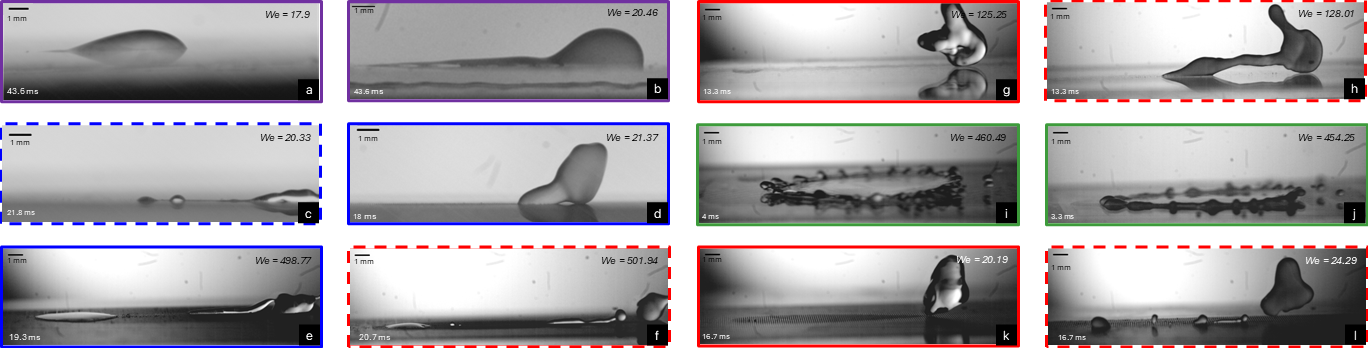}
        \caption{Representative high-speed images illustrating the influence of fluid composition, surface morphology, and wettability on droplet impact dynamics. (a,b) Deposition on uncoated glass for (a) water and (b) the emulsion. (c--f) Impact on the M50 microstructured surface: (c,e) water exhibiting partial sliding at low and high Weber numbers, respectively; (d) emulsion showing complete sliding at low Weber number; and (f) emulsion undergoing partial rebound at high Weber number. (g,h) Impact on Glaco-coated glass: (g) complete rebound of water and (h) partial rebound of the emulsion. (i,j) Splashing on Glaco-coated glass for (i) water and (j) the emulsion. (k,l) Impact on Glaco-coated M100 (k) complete rebound of water and (l) partial rebound of the emulsion. In all cases, the emulsion exhibits greater liquid retention, more persistent liquid bridges, and enhanced filament formation compared with water under similar impact conditions.}
    \label{fig:comparison}
\end{figure}

Comparing water and emulsion droplets under identical impact conditions reveals that fluid composition fundamentally modifies the impact dynamics and regime transitions (Fig.~\ref{fig:comparison}). Representative image sequences for the most relevant impact behaviours are shown in Fig.~\ref{fig:comparison}, highlighting the differences in spreading, retraction, and breakup between the two fluids. On uncoated glass Fig.~\ref{fig:comparison}a,b), water deposited only at the lowest Weber number ($We \approx 20$ before transitioning to sliding, whereas the emulsion remained in the deposition regime over a considerably wider range. The image sequences show that the emulsion spreads less extensively and retains a larger contact area throughout the impact, indicating greater viscous dissipation than water. This behaviour is consistent with previous studies showing that increasing liquid viscosity suppresses lamella expansion and enhances viscous energy dissipation, thereby delaying transitions toward rebound or splashing \cite{josserand2016drop, wang2021dynamics}. Similar viscosity-induced reductions in spreading have also been reported for complex fluids and concentrated suspensions \cite{piskunov2025spreading}. Nevertheless, the emulsion cannot be interpreted solely as a high-viscosity Newtonian liquid, since the dispersed oil droplets continuously modify the local interfacial properties during impact.

The different dynamics on the M50 surface are illustrated in Fig.~\ref{fig:comparison}c--f. Water develops a long, thin trailing film while remaining in partial contact with the substrate (Fig.~\ref{fig:comparison}c,e), whereas the emulsion undergoes complete downstream sliding at $We \approx 20$ (Fig.~\ref{fig:comparison}d). At $We \approx 500$, the emulsion begins to detach while remaining connected to the surface by a liquid bridge (Fig.~\ref{fig:comparison}f), signalling the transition toward partial rebound. Although increased viscosity would normally suppress droplet mobility, the opposite trend was observed. This counterintuitive behaviour suggests that the dispersed silicone oil forms a transient lubricating layer on the oleophilic microstructures, reducing contact-line pinning and promoting tangential motion. A similar lubrication mechanism was previously reported during emulsion impact on textured substrates, where oil redistribution altered the local wetting conditions and reduced adhesion \cite{quetzeri2024droplet}. Related studies on lubricant-infused and oil-covered textured surfaces have likewise demonstrated that thin liquid films can dramatically decrease contact angle hysteresis and facilitate droplet mobility \cite{wong2011bioinspired, Smith2013Droplet}.

The influence of the microstructure became even more evident on the M100 surface. Both liquids transitioned toward partial rebound for $We \gtrsim 30$, indicating that the larger interpillar spacing further reduced solid--liquid contact and favoured inertially driven recoil. However, the rebound morphologies differed substantially. Water formed relatively smooth elongated ligaments before detachment, whereas the emulsion generated irregular lobes connected by fragmented threads. These observations suggest that the dispersed oil phase redistributes kinetic energy during recoil and destabilises the retracting liquid sheet. Similar filamentary breakup has been observed during impacts of multiphase liquids and particle-laden drops, where local heterogeneities promote capillary instabilities and non-uniform ligament thinning \cite{damak2022dynamics, abbot2025elucidating}.

The effect of the superhydrophobic coating is evident in Fig. \ref{fig:comparison}g,h. Water recoils into a compact droplet that cleanly detaches from the substrate (Fig. \ref{fig:comparison}g), whereas the emulsion leaves behind a persistent liquid bridge that delays complete detachment (Fig. \ref{fig:comparison}h). This behaviour is consistent with numerous studies demonstrating that superhydrophobic coatings minimise contact time by reducing adhesion and viscous dissipation \cite{richard2002contact, bartolo2006bouncing, dwivedi2024dynamics}. The partial rebound present in the emulsion indicates that the oil phase locally anchors portions of the liquid despite the globally low surface energy, producing heterogeneous wetting conditions that are absent for pure water \cite{damak2022dynamics, quetzeri2024droplet}. At $We \approx 450$ (Fig. \ref{fig:comparison}i,j), both fluids splash; however, the emulsion produces thicker ligaments and a greater amount of residual liquid attached to the surface than water, indicating stronger local energy dissipation during lamella breakup.

On coated M50, both liquids completely rebounded at low Weber numbers. However, for $We \gtrsim 300$, their behaviour diverged considerably. Water continued to rebound cleanly, whereas the emulsion transitioned to splashing while maintaining localized contact with the substrate. This observation suggests that although the superhydrophobic coating reduces global adhesion, the multiphase structure of the emulsion promotes local stress concentrations that trigger ligament formation and breakup. Similar reductions of the splashing threshold on highly non-wetting surfaces have been attributed to the rapid expansion of the lamella and enhanced rim instabilities \cite{quintero2019splashing, xia2026droplet}.

The largest differences were observed on coated M100. Water exhibited complete rebound for $20 \lesssim We \lesssim 300$ (Fig. \ref{fig:comparison}), transitioning to splashing only at higher Weber numbers. In contrast, the emulsion already showed partial rebound with filament formation at $We \approx 20$ (Fig. \ref{fig:comparison}) and evolved into vigorous splashing with multiple fragmenting ligaments for $We \geq 270$. The earlier onset of fragmentation indicates that increasing pillar spacing destabilizes the expanding lamella more strongly for emulsions than for single-phase liquids. Previous investigations of textured surfaces have similarly reported that reduced pillar density lowers the resistance to liquid penetration and enhances capillary-driven breakup of the lamella \cite{zhang2018tunable, shi2025aerodynamic, cao2025dual, quetzeri2024droplet}. 

Overall, the experiments demonstrate that the impact dynamics of emulsions cannot be predicted solely from conventional dimensionless groups such as the Weber and Reynolds numbers. While inertia and surface topology determine the global impact regime, the dispersed oil phase continuously modifies local wetting, lubrication, and energy dissipation during impact.

\subsection{Wetted area after impact}

While the impact regimes provide a qualitative description of droplet behaviour, the residual wetted area $A_c$ offers a quantitative measure of liquid retention after impact. This quantity is particularly relevant for applications requiring efficient surface coverage, such as agricultural spraying and coating technologies. In the following, $A_c$ is normalised by the projected area of the incident droplet $A_p$, allowing the effects of fluid properties, surface morphology, and wettability to be compared directly.

\begin{figure}[htbp]
        \centering
        \includegraphics[width=1\textwidth]{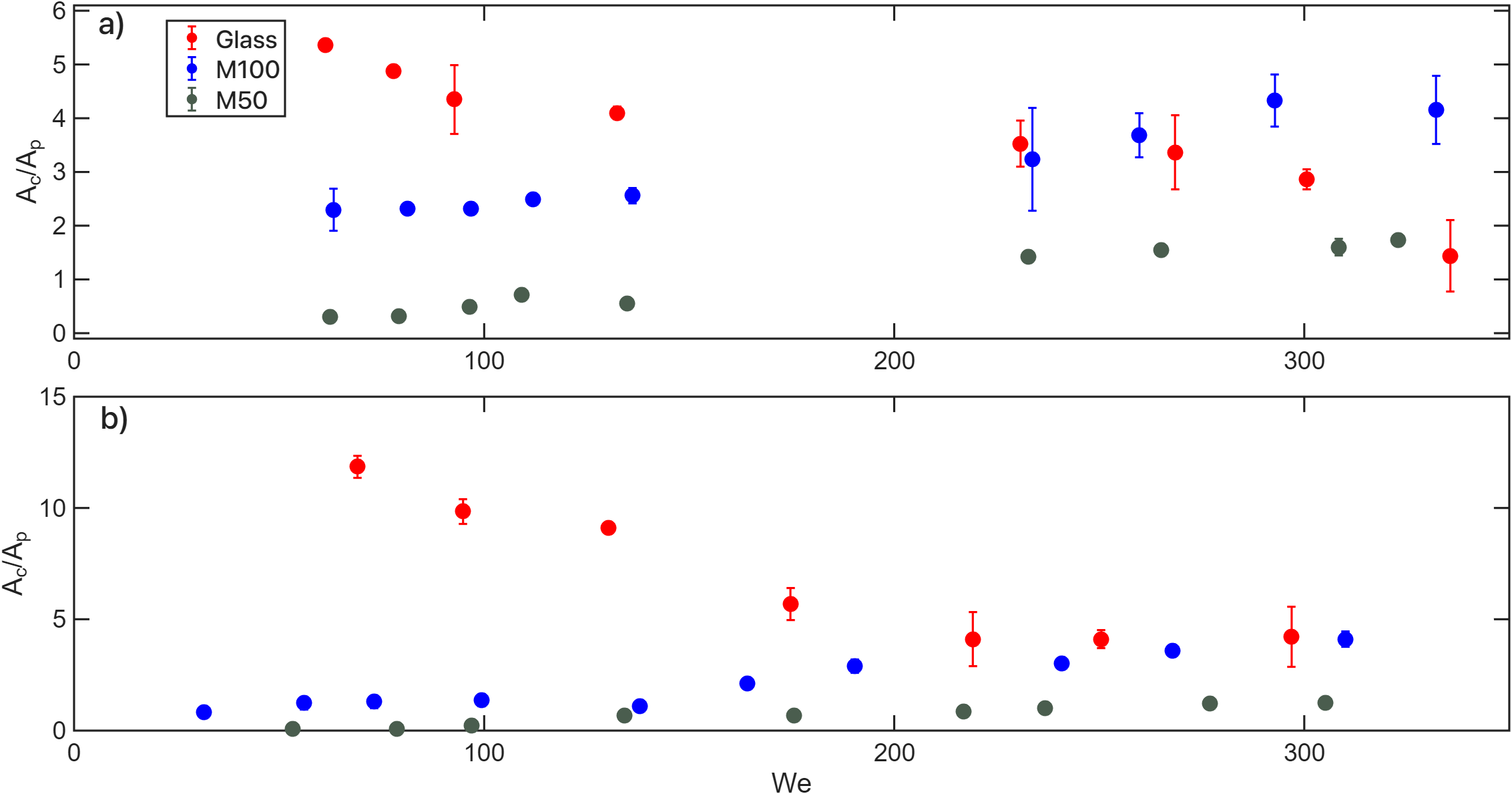}
        \caption{Normalised residual wetted area ($A_c/A_p$) as a function of Weber number for (a) water and (b) silicone oil-in-water emulsion droplets impacting inclined surfaces. Results are shown for the smooth glass substrate and the two microstructured surfaces (M50 and M100). Solid lines are included to guide the eye. The evolution of the wetted area reflects the competition between droplet spreading, recoil, liquid retention, and penetration into the surface texture.}
    \label{fig:coveredarea}
\end{figure}

For uncoated glass, the normalised wetted area decreased monotonically with increasing Weber number (Fig.~\ref{fig:coveredarea}a). This trend directly reflects the transition from deposition at low Weber numbers to sliding at higher impact velocities. As inertia increases, a larger fraction of the liquid is transported downstream before the droplet either detaches or exits the field of view, leaving progressively less liquid deposited on the surface. Similar reductions in residual liquid coverage with increasing impact velocity have been reported for water droplets impacting smooth inclined substrates, where inertial transport progressively dominates over capillary retention \cite{wang2021dynamics,sahoo2019interplay}.

In contrast, both microstructured surfaces exhibited an increase in wetted area with Weber number. Although the regime maps showed greater droplet mobility on M100 than on M50, the residual wetted area was consistently larger on M100 throughout most of the explored range. While the larger interpillar spacing promotes tangential motion by reducing contact-line pinning, it also lowers the pressure required for liquid penetration into the texture. Consequently, increasing impact pressure progressively drives a transition from a Cassie-Baxter state toward a partial Wenzel state, increasing the amount of liquid trapped within the microstructure after impact \cite{lee2010wetting,bartolo2006bouncing,reyssat2008impalement}. Similar transitions between suspended and impregnated wetting states have been observed on textured superhydrophobic surfaces subjected to increasing dynamic pressure \cite{lafuma2003superhydrophobic,krupenkin2004rolling}.

The emulsion displayed qualitatively similar trends (Fig.~\ref{fig:coveredarea}b). On glass, the wetted area decreased with Weber number, whereas on both microstructured surfaces the retained area increased with impact velocity. However, the absolute magnitude of the deposited area differed substantially from that of water.

On glass, emulsion droplets consistently covered a larger area than water droplets over the entire Weber number range. Unlike water, the emulsion leaves behind a thin residual oil film during spreading and recoil because the dispersed silicone oil possesses a greater viscosity and stronger affinity for the substrate than the surrounding aqueous phase. Similar behaviour has been reported during the impact of oil-in-water emulsions, where phase separation occurring during rapid spreading produces persistent oil deposits after the aqueous phase retracts \cite{damak2022dynamics,piskunov2025spreading}. Since the oil droplets experience lower capillary restoring forces than the continuous phase, they preferentially remain attached to the surface, increasing the apparent wetted area.

\begin{figure}
    \centering
    \includegraphics[width=1\linewidth]{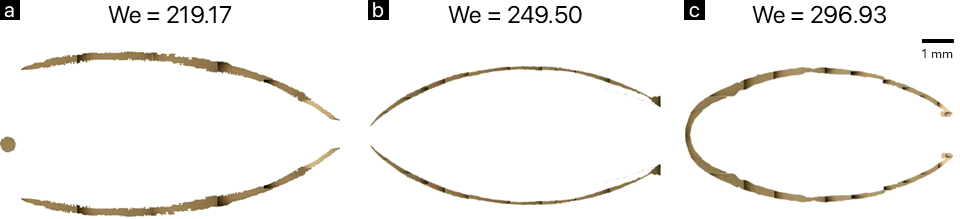}
    \caption{Microscopy reconstructions of oil deposits remaining on the glass substrate after impact of silicone oil-in-water emulsion droplets at different Weber numbers: (a) $We=219.17$, (b) $We=249.50$, and (c) $We=296.93$. The dispersed oil forms a characteristic elliptical ring-like deposit, with the size of the residual stain decreasing as the Weber number increases.}
    \label{fig:elipseshuecas}
\end{figure}

For $We=200-275$, $A_c/A_p$ of the emulsion was approximately four times larger than that of water. Optical microscopy revealed that the deposited stain consisted of an irregular elliptical ring with oil preferentially accumulated near the perimeter rather than uniformly distributed over the impact footprint (Fig.~\ref{fig:elipseshuecas}). Similar annular deposits have been reported for water-oil emulsions following impact and evaporation, where the rapid radial flow generated during lamella expansion transports dispersed oil droplets toward the contact line. These droplets subsequently coalesce and become pinned, forming an oil-rich peripheral ring \cite{vignes1996study}. Related segregation mechanisms have also been observed in evaporating emulsion droplets, where capillary flow and interfacial stresses drive preferential accumulation of the dispersed phase near the contact line \cite{deegan1997capillary}.

The fluid-free central region is likewise consistent with the presence of a transient air cushion beneath the impacting droplet. High-speed interferometric measurements have demonstrated that an air film forms immediately before contact, delaying wetting of the central region and redirecting the liquid radially outward \cite{thoroddsen2005air,driscoll2011ultrafast,kolinski2012skating}. The combination of this air-cushioning mechanism with outward transport of the dispersed oil phase provides a plausible explanation for the hollow elliptical deposits observed in the present experiments.

\begin{figure}
    \centering
    \includegraphics[width=1\linewidth]{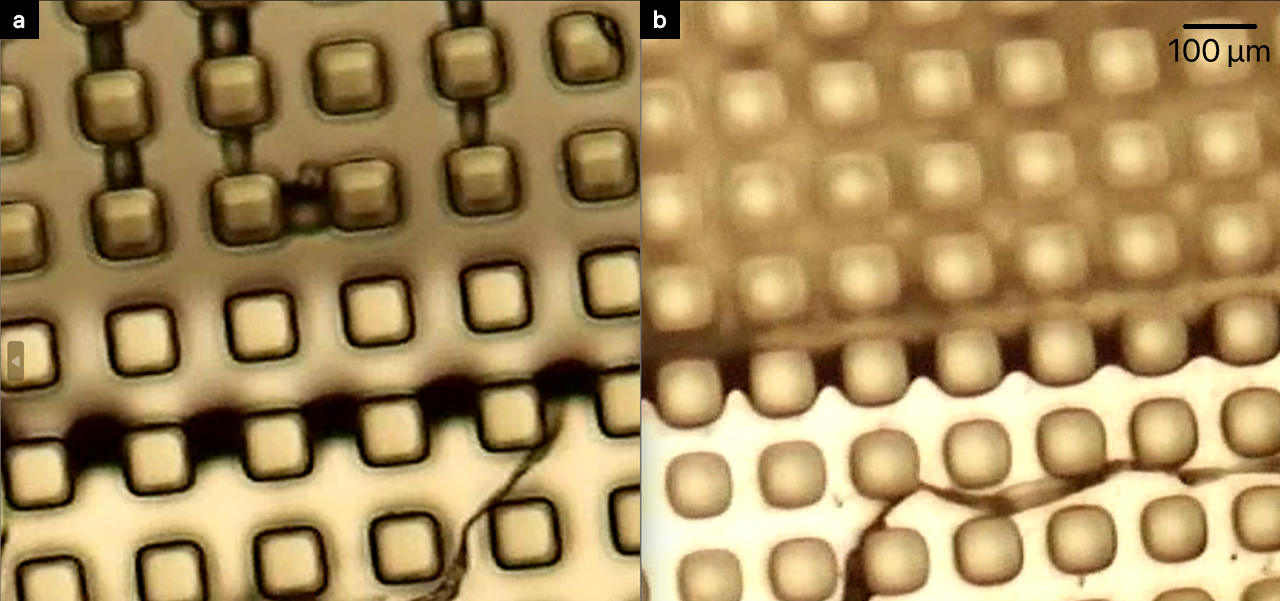}
    \caption{Optical microscopy images of liquid retained between the pillars of the M50 microstructured surface after droplet impact: (a) water and (b) silicone oil-in-water emulsion. The images reveal differences in the capillary bridges formed between neighbouring pillars and in the resulting liquid retention within the microstructure.}
    \label{fig:meniscos}
\end{figure}

For textured substrates water generally produced a larger wetted area than the emulsion. For M50, both fluids exhibited similar increases in covered area with Weber number, except for a localised deviation near $We\approx138$, where the emulsion spread slightly further (Fig.\ref{fig:coveredarea}b). Microscopic observations revealed distinct meniscus geometries between neighbouring pillars (Fig. \ref{fig:meniscos}). The emulsion formed capillary bridges with a smaller radius of curvature $R_c(\text{emulsion}) \approx 6.5$ $\mu$m than water $R_c(\text{water}) \approx 9.3$ $\mu$m, indicating stronger capillary suction within the texture. Previous studies of capillary bridges between microstructures have shown that increasing capillary pressure enhances liquid retention inside the texture while simultaneously limiting lateral spreading across the surface \cite{bouloudenine2026experimental}. Consequently, although the emulsion remains more strongly anchored within the microstructure, it covers a smaller projected area than water.

The same qualitative behaviour was observed on M100, where the emulsion wetted a smaller area than water over most of the explored Weber number range. Nevertheless, for $We\gtrsim200$, the two curves gradually converged. At these higher impact energies, the larger interpillar spacing facilitates both air evacuation and liquid penetration into the texture, reducing the influence of capillary pinning and allowing inertia to dominate the final wetted area. Similar inertia-driven reductions in the influence of surface texture have been reported during droplet impacts on structured substrates at high Weber numbers \cite{josserand2016drop}.

Overall, these results demonstrate that the final wetted area is determined not only by the impact regime but also by the interaction between impact pressure and surface microstructure. On smooth surfaces, increasing inertia primarily promotes downstream transport, thereby reducing liquid retention. On textured substrates, however, higher impact pressures increasingly force liquid into the microstructure, producing larger residual wetted areas despite greater droplet mobility. The dispersed oil phase further modifies this balance through lubrication, selective deposition, and enhanced capillary retention, highlighting that the final coverage of emulsion droplets cannot be inferred solely from their macroscopic impact behaviour.

\section{Conclusions}

The impact dynamics of silicone oil-in-water emulsion droplets on inclined microstructured surfaces were experimentally investigated and compared with those of water over a wide range of Weber numbers. The results demonstrate that the combined effects of fluid composition, surface morphology, and wettability govern not only the transient impact dynamics but also the final liquid retention on the substrate.

Five distinct impact regimes were identified: deposition, sliding, partial rebound, complete rebound, and splashing. Surface inclination introduced a pronounced asymmetry in the impact dynamics by promoting downstream spreading and shifting the balance between inertial, capillary, viscous, and adhesive forces. While smooth glass primarily favoured deposition and sliding, microstructured surfaces promoted rebound owing to the reduced liquid--solid contact area provided by the trapped air between the pillars. Increasing the interpillar spacing from 50 to 100 $\mu$m further enhanced droplet mobility by reducing contact-line pinning, although it also facilitated liquid penetration into the texture at higher Weber numbers.

The behaviour of the emulsion differed markedly from that of water. On smooth substrates, the emulsion remained in the deposition regime over a broader Weber number range and produced substantially larger residual wetted areas because the dispersed oil phase left persistent liquid films after impact. On microstructured surfaces, however, the emulsion exhibited enhanced sliding at low Weber numbers, suggesting that oil redistribution locally lubricated the texture and reduced contact-line pinning. At larger Weber numbers, the emulsion generated irregular ligaments, persistent liquid bridges, and earlier secondary breakup than water, indicating that the dispersed phase fundamentally modifies the redistribution of energy during droplet recoil and fragmentation.

Applying a superhydrophobic coating substantially reduced adhesion and promoted complete rebound for water over a broad range of impact conditions. In contrast, emulsion droplets frequently exhibited only partial rebound, leaving residual liquid connected by thin filaments even on highly non-wetting surfaces. These observations demonstrate that increasing surface hydrophobicity alone is insufficient to predict the impact behaviour of multiphase liquids, as local phase segregation and heterogeneous wetting introduce additional mechanisms absent in single-phase fluids.

Analysis of the residual wetted area further revealed that droplet mobility and liquid retention are not directly correlated. While larger pillar spacing promoted downstream motion, it also increased liquid penetration into the microstructure, resulting in larger deposited areas at high Weber numbers. On smooth substrates, the emulsion consistently deposited more liquid than water, whereas on textured surfaces stronger capillary confinement within the microstructure limited the projected wetted area despite enhanced liquid retention inside the texture.

Overall, this work demonstrates that the impact dynamics of emulsions cannot be predicted solely from conventional dimensionless groups such as the Weber and Reynolds numbers. The dispersed oil phase continuously modifies local wetting, lubrication, and energy dissipation throughout the impact process, giving rise to impact regimes and deposition patterns that differ fundamentally from those of homogeneous liquids. These findings provide new insight into the interaction between complex fluids and inclined textured surfaces and may contribute to the design of more efficient spray deposition processes in applications ranging from agricultural pesticide delivery to functional coatings and thermal management.

\section*{Acknowledgments}
We thank Maricela Zapata Arroyo for her help with the rheological measurements. We thank Carina Granados for her assistance in fabricating the microstructured surfaces. We thank Carlos Palacios Morales for facilitating the high-speed camera for the experiments. M.A.Q-S. acknowledges support from DGAPA through the grant PAPIIT-UNAM IA101025.

\bibliographystyle{elsarticle-num} 
\bibliography{bibliography}

\end{document}